\documentclass[twocolumn,prd,nofootinbib,aps,prl,floats,floatfix,amsmath,amssymb,longbibliography,secnumarab
ic]{revtex4-1} %
\usepackage[final]{graphicx}
\usepackage{hyperref}
\usepackage{amsmath}
\usepackage{bbm}
\usepackage{amsfonts}
\usepackage{amssymb}
\usepackage{latexsym}
\usepackage{graphicx}
\usepackage[english]{babel}
\usepackage{multirow}
\usepackage{float}
\usepackage{url}
\usepackage{slashed}
\usepackage{xcolor} 
\usepackage[utf8]{inputenc}

\def\sfrac#1#2{{\textstyle{#1\over #2}}}
\newcommand{\be}{\begin{equation}}
\newcommand{\ee}{\end{equation}}
\newcommand{\ba}{\begin{array}}
\newcommand{\ea}{\end{array}}
\newcommand{\bea}{\begin{eqnarray}}
\newcommand{\eea}{\end{eqnarray}}
\newcommand{\sss}{\scriptscriptstyle}

\newcommand{\nn}{\nonumber}

\newcommand{\R}{{\sss R}}

\begin{document}

\title{Affleck-Dine inflation}

\author{James M.\ Cline}
\author{Matteo Puel}
\author{Takashi Toma}
\affiliation{McGill University, Department of Physics, 3600 University St.,
Montr\'eal, QC H3A2T8 Canada}
\begin{abstract}
The Affleck-Dine mechanism in its simplest form provides baryogenesis
from the out-of-equilibrium evolution of a complex scalar field with a
simple renormalizable potential.  We show that such a model, supplemented
by nonminimal coupling to gravity, can also provide inflation, consistent
with Planck constraints, simultaneously with the generation of the baryon
asymmetry.  The predictions of the model  include significant 
tensor-to-scalar ratio and possibly baryon isocurvature fluctuations. 
The reheating temperature is calculable, making the model fully
predictive. We require color triplet scalars for reheating and
transfering the primordial baryon asymmetry to quarks; these could be
observable at colliders.  They can also be probed at higher scales by
searches for quark compositeness in dijet angular distributions, and
flavor-changing neutral current effects.
\end{abstract}
\maketitle

\section{Introduction}

Theoretical mechanisms for baryogenesis abound and take many very
different forms, but one common attribute is that they occur at some
cosmological epoch following inflation.  This seems like a necessity,
since exponential expansion should dilute any preexisting baryon
asymmetry.  In this work we present an exception, showing that it is
possible to generate the baryon asymmetry of the universe (BAU) during the course of
inflation, using the inflaton as an essential element.

One of the earliest proposed baryogenesis mechanisms was that
of Affleck and Dine (AD) \cite{Affleck:1984fy}  in which a complex scalar field carrying
baryon number can spontaneously create the BAU starting from field
values displaced from the minimum of the potential.  A
baryon-violating coupling is required to satisfy Sakharov's 
requirements \cite{Sakharov:1967dj}.  
Although the AD mechanism is most commonly implemented in
supersymmetric models whose potentials have nearly flat directions, 
it was originally demonstrated using a simple renormalizable potential of the 
form
\be
	V_J = m_\phi^2|\phi|^2 + \lambda|\phi|^4 +
i\lambda'(\phi^4-\phi^{*4})
\label{potential}
\ee
in the seminal reference \cite{Affleck:1984fy}.  (We
use the subscript to denote the Jordan frame since a change of
frames will be invoked below.)  When $\lambda'=0$,  the 
potential has a U(1) global symmetry, that we will identify with baryon number.
A generic initial condition such that $\langle\phi\rangle\ne 0$
spontaneously breaks CP and thermal equilibrium, as also required by Sakharov.  The field
winds around, at first generating baryon number, until Hubble damping
of $\langle\phi(t)\rangle$ makes the $\lambda'$ interaction
negligible, and baryon number becomes conserved.

The same kind of potential could be used for a two-field version of
chaotic inflation \cite{Linde:1983gd}.  Constraints from the Planck
experiment now disfavor  chaotic inflation with $\phi^2$ or $\phi^4$
potentials \cite{Akrami:2018odb} since they predict too high
tensor-to-scalar ratio $r$, given the measured value of the scalar
perturbation spectral index $n_s = 0.965\pm 0.004$.  However this
problem can be cured by adding a nonminimal coupling to gravity (we
write $2\xi$ rather than $\xi$ to agree with the usual convention for
inflation along a single component of the complex scalar),
\be
	{\cal L}_{J} = {m_P^2\over 2}R \left(1 +
	2\xi|\phi|^2\right)
\label{nmcoup}
\ee
where $m_P = 2.44\times 10^{18}\,$GeV is the reduced Planck mass,
that we set to 1 unless explicitly shown. 
This introduces a noncanonical kinetic
term for $\phi$ upon Weyl-transforming to the Einstein
frame, and it flattens the potential at large field values to make
the predictions of the model compatible with Planck observations,
in the case of a real scalar field inflaton 
\cite{Okada:2010jf,Linde:2011nh}.  Our goal is to determine whether
this can still hold true for the two-field model, while at the same
time generating the observed baryon asymmetry.  A potential issue
is that isocurvature perturbations can be produced in two-field
models, and these are constrained by the Planck observations.

A similar idea was explored in ref.\ \cite{Evans:2015mta}, using the
complex sneutrino field in a supergravity model as both the inflaton
and the AD field for leptogenesis.  The model is rather special and
intricate, whereas ours could be considered simpler and more generic.
The predictions of $r$ versus $n_s$ illustrated there are in
significant tension with the latest data.   Moreover isocurvature
fluctuations were not considered in ref.\ \cite{Evans:2015mta}.  

\section{Model}
Eqs.\ (\ref{potential},\ref{nmcoup}) are sufficient to determine
the inflationary trajectory until the epoch of reheating.
It is convenient to make a Weyl
rescaling of the metric, $g_{\mu\nu} \to\Omega^2  g_{\mu\nu}$,
with $\Omega^2 = 1/(1 + 2\xi|\phi|^2)$.  The Lagrangian
in the Einstein frame, including gravity, is then
\be
	{\cal L}_{E}= \sfrac12 R + \Omega^4\left(
	{|\partial\phi|^2\over\Omega^2} + 
3\xi^2(\partial|\phi|^2)^2
	- V_J \right)\, .
\ee
Writing the complex scalar as $\phi = (X + iY)/\sqrt{2}$ and ignoring
spatial gradients, the scalar kinetic term takes the form
\be
	{{\cal L}_{\rm kin}} = \sfrac12\Omega^2(\dot X^2+\dot Y^2)
	+ 3\Omega^4\xi^2(X\dot X + Y\dot Y)^2  
\label{Lkin}
\ee
with $\Omega^2=1/(1 +\xi(X^2+Y^2))$.  Thus $X$ and $Y$ are not
canonically normalized fields.  Instead of reexpressing them in terms
of such fields, we will numerically solve the equations of motion for 
$X$ and $Y$ to determine the predictions for inflation and baryogenesis.
Details of deriving the first-order equations convenient for
numerical integration can be found for
example in ref.\ \cite{Cline:2006hu} (see eqs.\ (2.100-2.101)).
We choose initial conditions close to the inflationary attractor
solution, by setting the derivatives of the canonical momenta
$\Pi_{X} = d{\cal L}/d\dot X$, $\Pi_{Y} = d{\cal L}/d\dot Y$
initially to zero.

More is needed in order to get reheating and transfer of the baryon 
asymmetry, initially stored in $\phi$, into quarks.  A natural
option for reheating is the Higgs portal coupling $\lambda_{\phi h}
|\phi|^2|H|^2$.  However since we also need a coupling to quarks,
it is simpler to use the same interactions both for reheating and for
transfer of  the baryon asymmetry.  This can be accomplished by
introducing three QCD triplet scalars $\Phi_i$ carrying baryon number $2/3$,
with couplings
\bea
	V_\Phi &=& \epsilon_{abd}\left(\lambda'' \,\phi^*\, \Phi_1^a
\Phi_2^b\Phi_3^d + y_1\Phi_1^a\, \bar u_\R^b d_\R^{c,d} \right.\nn\\
&+& \left. y_2\Phi_2^a\, \bar u_\R^b d_\R^{c,d}  +  
y_3\Phi_3^a\, \bar d_\R^b d_\R^{c,d}\right) + {\rm H.c.}
\label{Phic}	
\eea 
where $a,b,d$ are color indices, the quarks are right-handed (SU(2)$_L$ singlets) and
$d_\R^c$ denotes the conjugate down-type quark.  For simplicity we omit
generation indices on the quarks and the Yukawa couplings $y_i$.
These interactions allow for the decay $\phi \to uudddd$ via 
virtual $\Phi_i$ exchange, 
and imply that $\phi$ carries baryon number 2.  The same conclusion 
holds if we choose $\Phi_1 \bar u_\R u_\R^c$ and $\Phi_{2,3}\bar d_\R
d_\R^c$
couplings instead of (\ref{Phic}).

For small values of the $\lambda''$ coupling, we can view reheating as
occurring through the perturbative decays $\phi\to
\Phi_1\,\Phi_2\,\Phi_3$, which rapidly thermalize with the quarks
and thereby the rest of the standard model degrees of freedom.
Assuming that $\phi$ is much heavier than $\Phi_i$, the decay rate is
\be
	\Gamma_\phi = {3\,\lambda''^2 \over 256\pi^3}\,m_\phi
\label{decayrate}
\ee

\section{Inflation $+$ baryogenesis}

An interesting aspect of our model is that the same parameters that
influence inflationary observables can also affect the magnitude of
the baryon asymmetry.  Thus, although we describe the two
processes separately, a fully viable model depends upon the
interplay between the two.

\begin{table*}[]
\centering
\tabcolsep 2.5pt
\begin{tabular}{|c|c|c|c|c|c|c|c|c|c|c|c|c|}
\hline
model & $m_\phi/m_P$ & $\lambda$ & $\lambda'$ & $\xi$ & $\lambda''$ & $X_0$ &
$Y_0$ & $N_{\rm tot}$ & $N_*$ & $n_s$ & $r$ & $T_{RS}$\\
\hline
1 &   $6.43\times 10^{-7}$ &   $8.73\times 10^{-12}$ &    
$6.89\times 10^{-13}$ &  $5.96\times 10^{-2}$  & 
$8.67\times 10^{-5}$ &   $18.4$ &  $6.63$  & $65$ & $53.1$  & $0.962$    &  
$1.4\times 10^{-2}$ &   $9\times 10^{-2}$\\    
2& $4.67\times 10^{-7}$ &   $3.49\times 10^{-11}$ &    $6.68\times
10^{-13}$ &   $0.180$ &       $2.93\times 10^{-5}$
&    $23.7$ &       $0.91$ &      $146$ &  $52.0$ &  $0.961$  &        
$7.2\times 10^{-3}$  &  $4\times 10^{-5}$ \\ 
\hline   
\end{tabular}
\caption{Parameters and initial values for two benchmark models,
including the total number of $e$-foldings of inflation $N_{\rm tot}$,
number of $e$-foldings before horizon crossing $N_*$, spectral
index $n_s$ (evaluated at $k_*=0.05$\,Mpc$^{-1}$), tensor-to-scalar ratio $r$ and off-diagonal transfer
matrix element $T_{RS}$, which is a measure of the correlation between
adiabatic and isocurvature perturbations.
\label{tab1}}
\end{table*}

\begin{figure*}[t]
\begin{center}
 \includegraphics[scale=0.4]{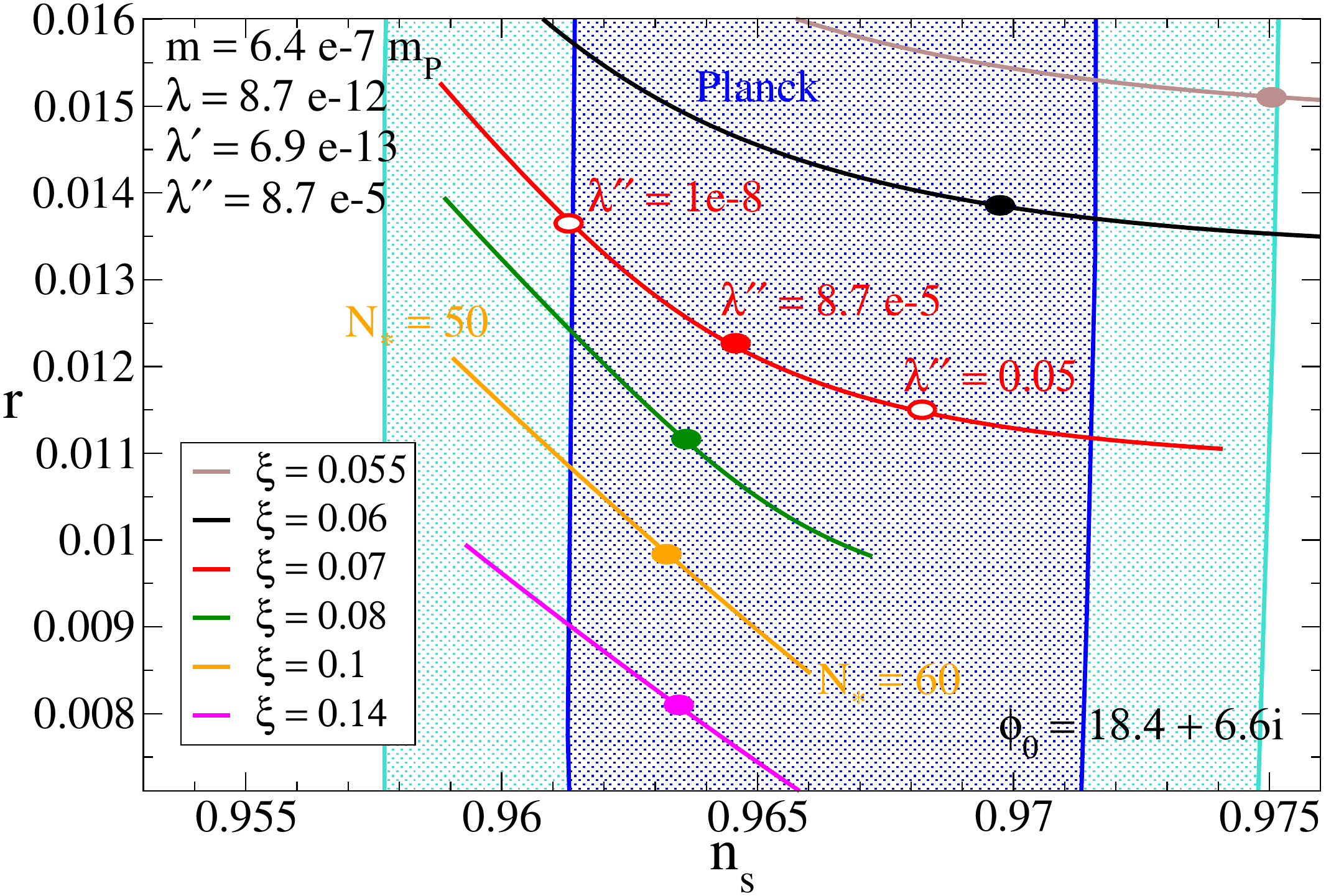}\hfil
\includegraphics[scale=0.4]{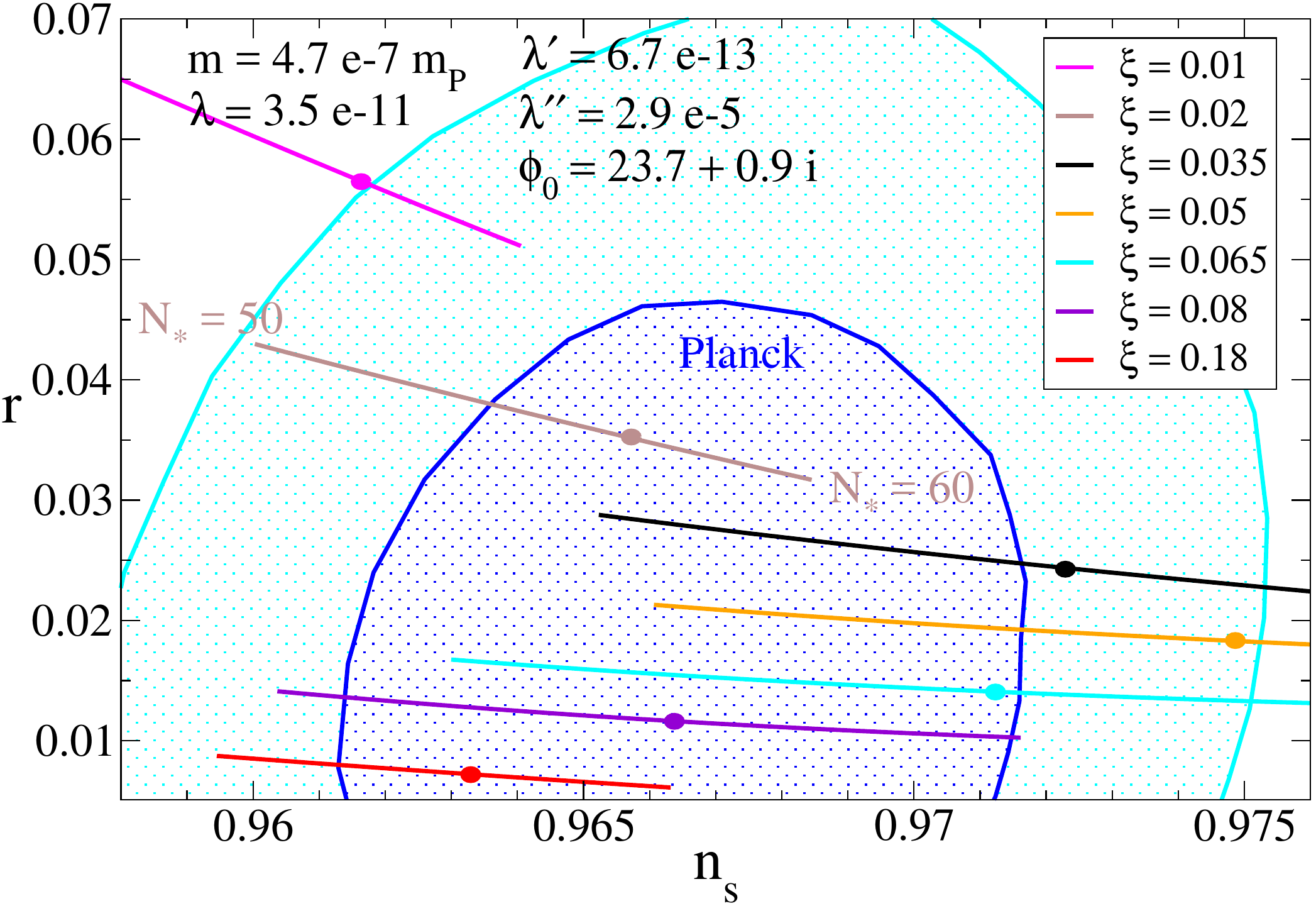}
 \caption{Scalar-to-tensor ratio versus spectral index for several
values of the nonminimal coupling $\xi$, varied around the parameters
of models 1 (left) and 2 (right) given in table \ref{tab1}.  The pivot
scale is $k_*=0.002$\,Mpc$^{-1}$ for comparison with the Planck
$1\,\sigma$ and $2\,\sigma$ allowed regions.  The number of
$e$-foldings between horizon crossing and the end of inflation, $N_*$,
is allowed to vary between 50 and 60, but the definite values shown by
the solid dots are predicted by making a specific choice of
$\lambda''$.  The dependence on $\lambda''$ is shown on the $\xi=0.07$
curve for model 1.}
 \label{fig:nsr}
\end{center} 
\end{figure*}

\begin{figure*}[t]
\begin{center}
 \includegraphics[scale=0.4]{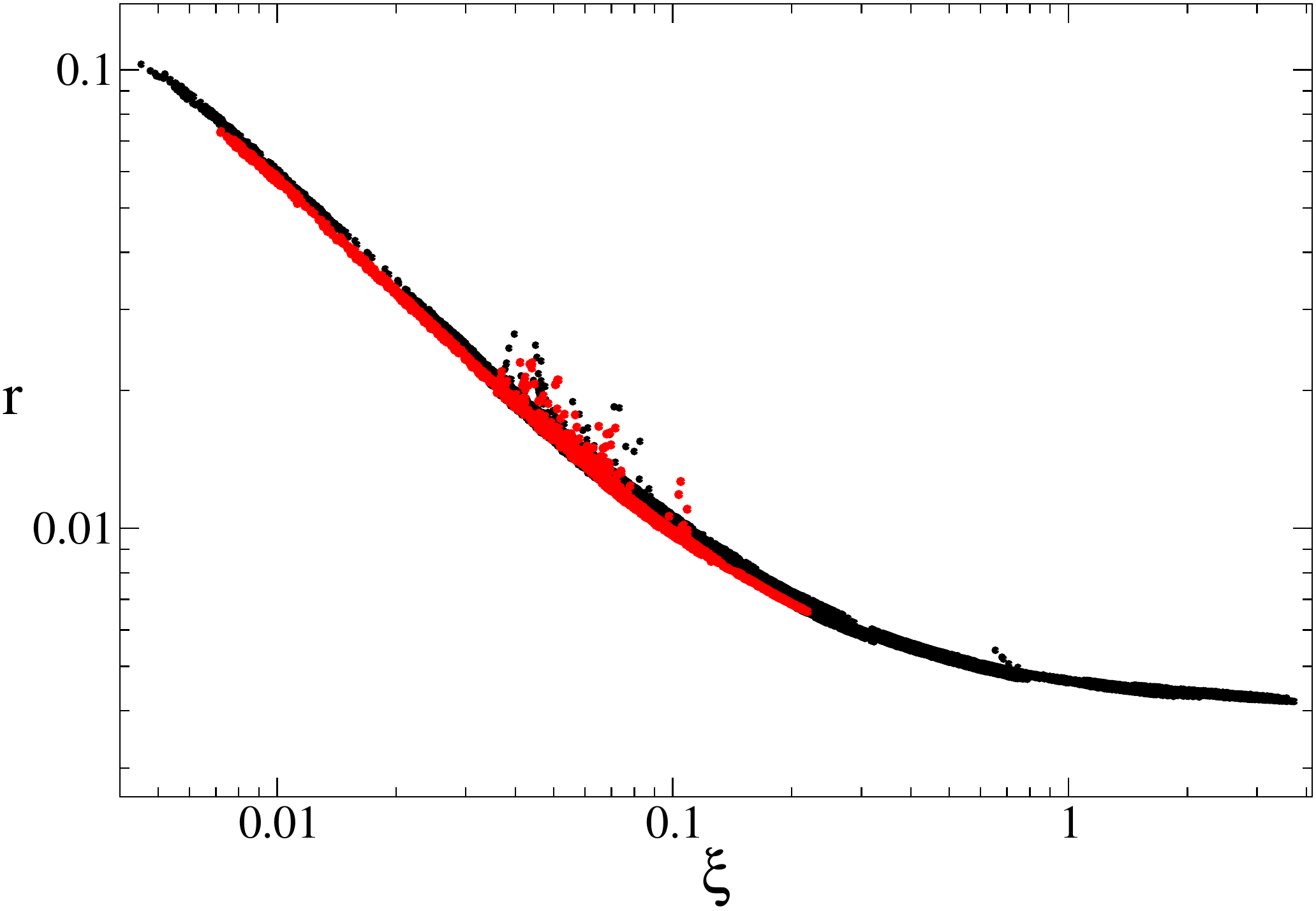}\hfil
\includegraphics[scale=0.4]{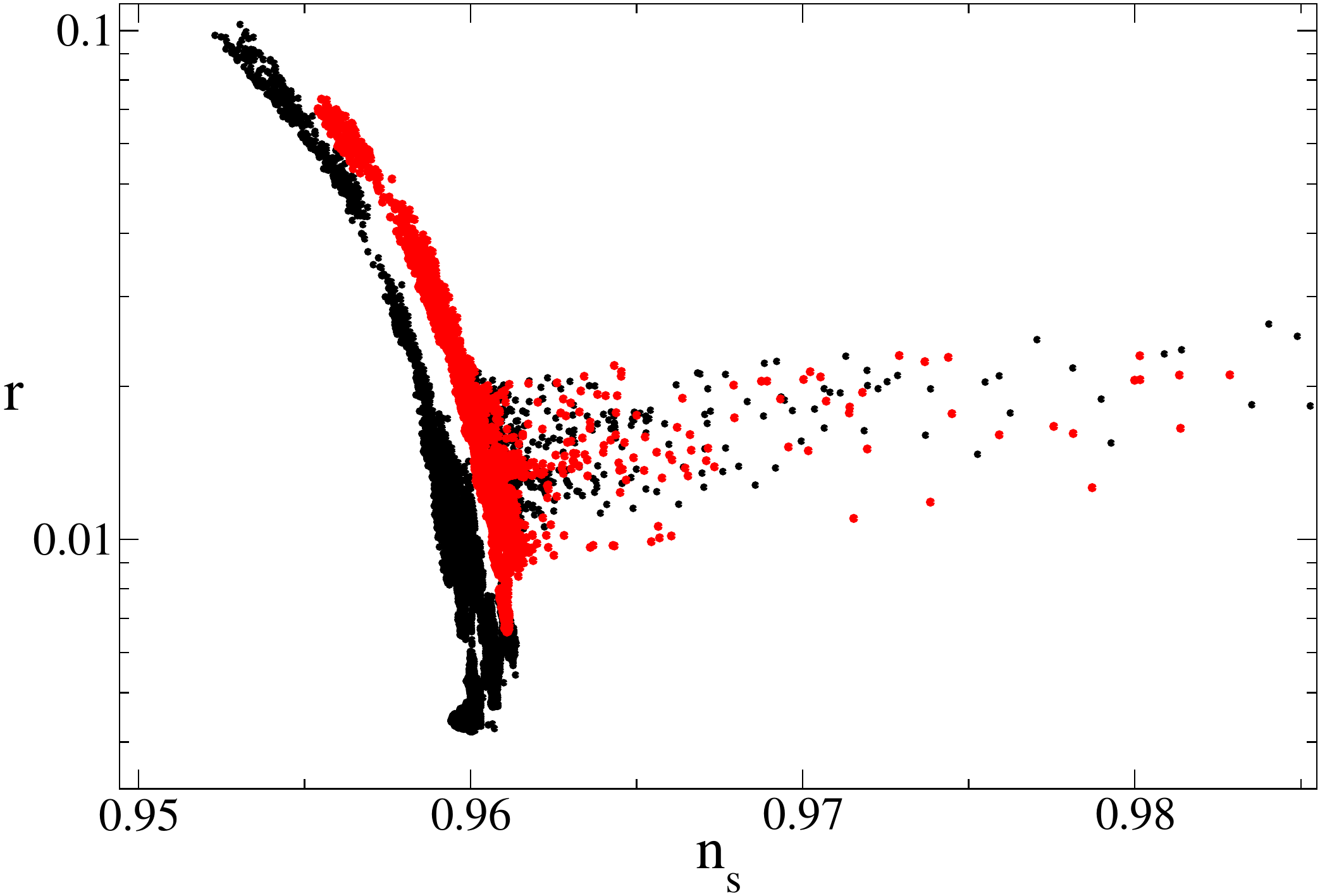}
 \caption{Scatter plots from the Markov Chain Monte Carlo (MCMC) search
of parameter space.  Left: correlation of $r$ with $\chi$.  Right:
correlation of $r$ (evaluated at $k_*=0.002\,$Mpc$^{-1}$) with $n_s$
(at $k_*=0.05\,$Mpc$^{-1}$).  Black versus red points correspond to
two different chains as described in the text.}
 \label{fig:mcmc}
\end{center} 
\end{figure*}

\subsection{Slow-roll parameters}

Although we can solve for the inflaton trajectories without reference
to the canonically normalized fields, that we will denote by 
$(U,V)$,
it is necessary to know them for computing inflationary
perturbations.  It is straightforward to diagonalize the kinetic term
(\ref{Lkin}) at a given point in field space to find
\bea
	\left(\dot X\atop\dot Y\right) &=& 
\left({c_\theta\atop s_\theta}\
	{-s_\theta \atop \phantom{-}c_\theta} 
\right)
\left( {e_1\atop 0}\ {0\atop e_2}\right)\left({c_\psi\atop -s_\psi}\
	{s_\psi \atop \phantom{-}c_\psi}\right)
	\left(\dot U\atop \dot V\right) \nn\\
 &\equiv& Z_0 R_\psi\left(\dot U\atop \dot V\right) \equiv  Z\left(\dot U\atop \dot V\right)
\label{diag_kin}
\eea
with $e_1 = \Omega^{-1}(1 +
6\Omega^2\xi^2(X^2+Y^2))^{-1/2}$, $e_2 = \Omega^{-1}$, 
$\theta  = \tan^{-1}(Y/X)$.
Then ${\cal L}_{\rm kin} = \sfrac12(\dot U^2 + \dot
V^2)$.  The matrix $Z$ allows us to transform slow roll parameters
computed in the original field basis (indices $i,j$) to those in 
the canonical basis (indices $m,n$):
\bea
	\epsilon_m &=& {\left(Z_{im}\,\partial_i V_E\right)^2\over 2
V_E^2} \nn\\
	\eta_{mn} &=& Z_{im}Z_{jn} {\partial_i\partial_j
	V_E\over V_E}
	+Z_{im}\partial_{i}Z_{jn}{\partial_{j}V_E\over V_E}
\label{srp}
\eea
where $V_E = \Omega^4 V_J$ is the Einstein frame potential.

The extra rotation $R_\psi$ in eq.\ (\ref{diag_kin}) is not necessary for
diagonalizing the kinetic term, but it is required in order to be able
to interpret $Z$ as the Jacobian matrix $\partial(X,Y)/\partial(U,V)$.
If we omit $R_\psi$ so that $Z=Z_0$, such an interpretation is not
generally consistent since then the relations
\bea
	U_{,YX} = \partial_X(Z_0^{-1})_{12}  &=&
	\partial_Y(Z_0^{-1})_{11} = U_{,XY}\nn\\
	V_{,YX} = \partial_X(Z_0^{-1})_{22} &=&
	\partial_Y(Z_0^{-1})_{21} = V_{,XY}
\label{jac_cond}
\eea
may not be satisfied.  
We are free to set $\psi=0$ at a given point in field space, such as
the point of horizon crossing, 
but not its derivatives.  Eqs.\ (\ref{jac_cond}) with $Z_0\to Z_0
R_\psi$ imply
\bea
	(Z_0^{-1})_{22}\,\psi_{,X} - (Z_0^{-1})_{21}\,\psi_{,Y}  &=&
	(Z_0^{-1})_{12,X} -(Z_0^{-1})_{11,Y}\nn\\
	-(Z_0^{-1})_{12}\,\psi_{,X} + (Z_0^{-1})_{11}\,\psi_{,Y} &=&
	(Z_0^{-1})_{22,X} - (Z_0^{-1})_{21,Y}\, .\nn
\eea
This has the solution
\bea
	\left(\psi_{,X}\atop\psi_{,Y}\right) &= &
	\det Z_0\,(Z_0^{-1})^T\left((Z_0^{-1})_{12,X} - (Z_0^{-1})_{11,Y}\atop
	(Z_0^{-1})_{22,X} - (Z_0^{-1})_{21,Y}\right)\nn\\
	&=& \Omega^2{e_1 e_2\over X^2+Y^2}\left( {-Y\atop \phantom{-}X} 
	\right)\, .
\eea
The consistent identification of $Z$ with a Jacobian matrix insures
that $\eta_{mn}$ is symmetric in $mn$, even though the second term
in (\ref{srp}) is not explicitly symmetric.  Then we can write the
second term in eq.\ (\ref{srp}) as
\be
	\partial_i Z = (\partial_i Z_0)\, R_\psi + \psi_{,i} Z_0 
\partial_\psi R_\psi \,.
\ee

To compute the adiabatic perturbation spectrum and the
tensor-to-scalar ratio, we use the slow-roll formalism of ref.\
\cite{Byrnes:2006fr}, evaluating the slow-roll parameters (\ref{srp})
along the numerically determined inflationary solutions.  This
requires going from the $U,V$ basis of the canonical fields to
the $\sigma,s$ basis of adiabatic/entropy directions, defined by
\bea
	d\sigma &=& c_\alpha dU + s_\alpha dV\nn\\
	ds &=& -s_\alpha dU + c_\alpha dV
\label{dseq}
\eea
with $\alpha = \tan^{-1}(\dot V/\dot U)$.  The rotated slow roll parameters
are given by \cite{Gordon:2000hv}
\bea
	\epsilon_\sigma &=& (c_\alpha \partial_U V_{E} + s_\alpha 
	\partial_V V_{E})^2/
(2 V_E^2)\nn\\
	\epsilon_s &\cong& 0\nn\\
	\eta_{\sigma\sigma} &=& c_\alpha^2 \eta_{UU} + 2 c_\alpha s_\alpha \eta_{UV}
	+ s_\alpha^2 \eta_{VV}\nn\\
	\eta_{ss} &=& s_\alpha^2 \eta_{UU} -2 c_\alpha s_\alpha \eta_{UV}
	+ c_\alpha^2 \eta_{VV}\nn\\
	\eta_{\sigma s} &=& c_\alpha s_\alpha (\eta_{VV} - \eta_{UU})
	+(c_\alpha^2 - s_\alpha^2)\eta_{UV}
\eea
Then to leading order in the slow-roll expansion,
the scalar spectral index and tensor-to-scalar ratio are \cite{Byrnes:2006fr}
\bea
	n_s &=& 1 -(6-4 c_\Delta^2)\epsilon_\sigma + 2 s_\Delta^2
\eta_{\sigma\sigma} + 4 s_\Delta c_\Delta \eta_{\sigma s} +
	2 c_\Delta^2 \eta_{ss}\nn\\
	r &=& 16\epsilon_\sigma
\eea
where $c_\Delta = -2{\cal C}\,\eta_{\sigma s}$, $s_\Delta =
+\sqrt{1-c_\Delta^2}$, ${\cal C} = 2-\ln 2-\gamma\cong 0.73$
($\gamma$ is the Euler constant),
and the derivatives
of $V_E$ with respect to $U$ and $V$ are computed similarly to
eq.\ (\ref{srp}).  
Including the effect of isocurvature modes ($T_{RS}$), which we will 
explain below, the scalar amplitude is
\be
	A_s = {V_*\over 24\pi^2\,\epsilon_\sigma}
\left[1 -2\epsilon_\sigma +
2{\cal C}\,(3\,\epsilon_\sigma-\eta_{\sigma\sigma} -2\eta_{\sigma s}T_{RS})\right]
\label{Aseq}
\ee
with $V_* = V_E$ evaluated at horizon crossing and we have neglected
terms of order $T_{RS}^2$.

We searched the parameter space via Markov Chain Monte Carlo (MCMC)
to find models in agreement with Planck constraints on $A_s$, $n_s$, 
$r$ and the baryon asymmetry (discussed below).  Two benchmark models
are identified in table \ref{tab1}.  
The correlation of $r$ with $n_s$
is shown over the interval $N_*=(50,60)$ $e$-foldings, for several
values of $\xi$ and fixed values of the potential parameters
corresponding to  the two benchmark models
in fig.\ \ref{fig:nsr}.  

On each curve a heavy dot is indicated to
show the prediction of the model, for the chosen value of $\lambda''$,
that determines the reheating temperature and thus the number of
$e$-foldings $N_*$ between horizon crossing and the end of inflation.
The value of $N_*$ is determined by solving eq.\ (47) of ref.\ \cite{Akrami:2018odb}
(see also ref.\ \cite{Liddle:2003as}), 
\be
	 N_* = 67 -\ln\left(k_*\over H_0\right) + \sfrac14\ln
	\left(V_*^2\over m_P^4\rho_{\rm end}\right) +
	\sfrac1{12}\ln\left(\rho_{\rm rh}\over g_* \rho_{\rm end}\right)
\label{Nsteq}
\ee
where $H_0$ is the Hubble constant today, $\rho_{\rm end}$ is the energy
density at the end of inflation, $g_*=106.75 + 18$ 
(counting the extra degrees of freedom from the
colored scalars), and the reference scale
$k_* = 0.002$\,Mpc$^{-1}$ for comparison with the Planck preferred
regions in the $n_s$-$r$ plane.   The energy density at the time of
reheating is $\rho_{\rm rh} = \sfrac43\Gamma_\phi^2 m_P^2$, as explained
below---see eq.\ (\ref{rho_rh}); this makes $N_*$ depend upon $\lambda''$ as
$N_*\sim \sfrac13\ln\lambda''.$   Since $V_*$ appears in eq.\
(\ref{Nsteq}) but also depends upon $N_*$, we rescale the parameters of the potential while
iteratively determining $N_*$, keeping $A_s = e^{3.044}\times
10^{-10}$ fixed to the observed central value \cite{Akrami:2018odb}.  To illustrate the
dependence on $\lambda''$, we indicate two other horizon-crossing
positions on the $\xi=0.07$ curve of model 1, for larger and smaller
values of $\lambda''$.  The relation between $\lambda''$ and the
reheat temperature will be discussed in section \ref{bgrh}.

The strong correlation between the tensor ratio $r$ and the 
nonminimal coupling $\xi$ is also clearly seen in the larger sample of
models from two MCMC chains, fig.\ \ref{fig:mcmc} (left).  The points shown
have a total $\chi^2 < 10$, defining $\chi^2$ in the usual way in terms
of the observables $r$, $n_s$ and $\eta_B$,
\be
	\chi^2 = \sum_i {(x_i -\bar x_i)^2\over \delta x_i^2}
\ee
summed over observables $x_i$ with central value $\bar x_i$ and
experimental error $\delta x_i$.
The black points come from
a chain where the experimental limit on $r$ was somewhat relaxed.
The correlation between $r$ and $n_s$ within the chains is also notable,
as shown in fig.\ \ref{fig:mcmc} (right).  In both plots, one can notice
a population of models scattered away from the main trends.  These are
special cases in which the total number of $e$-foldings of inflation are
not much greater than the minimum required, $N_e\sim 60$.  We will
discuss these cases in more detail below.

\begin{figure*}[t]
\begin{center}
 \includegraphics[scale=0.4]{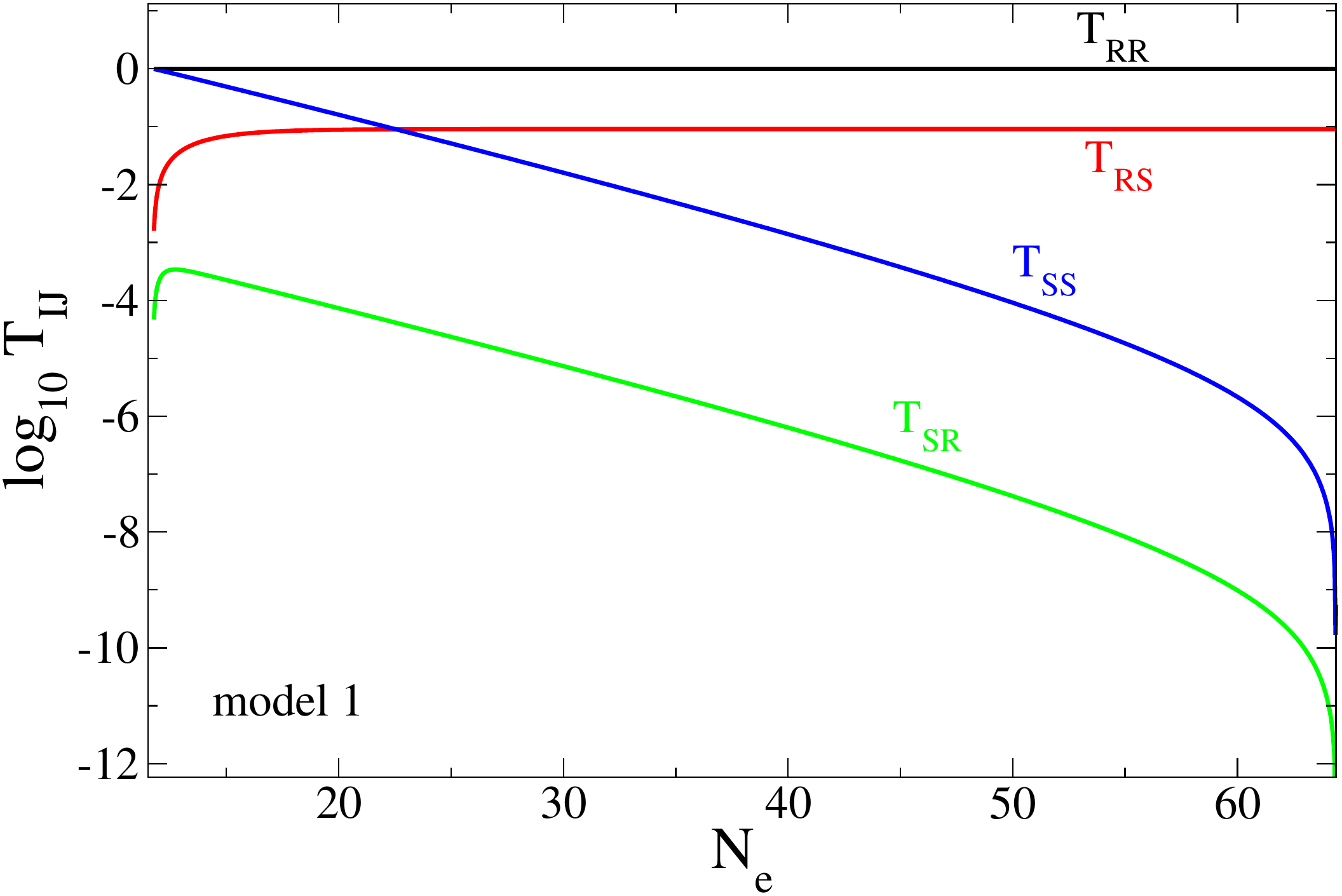}\hfil
\includegraphics[scale=0.4]{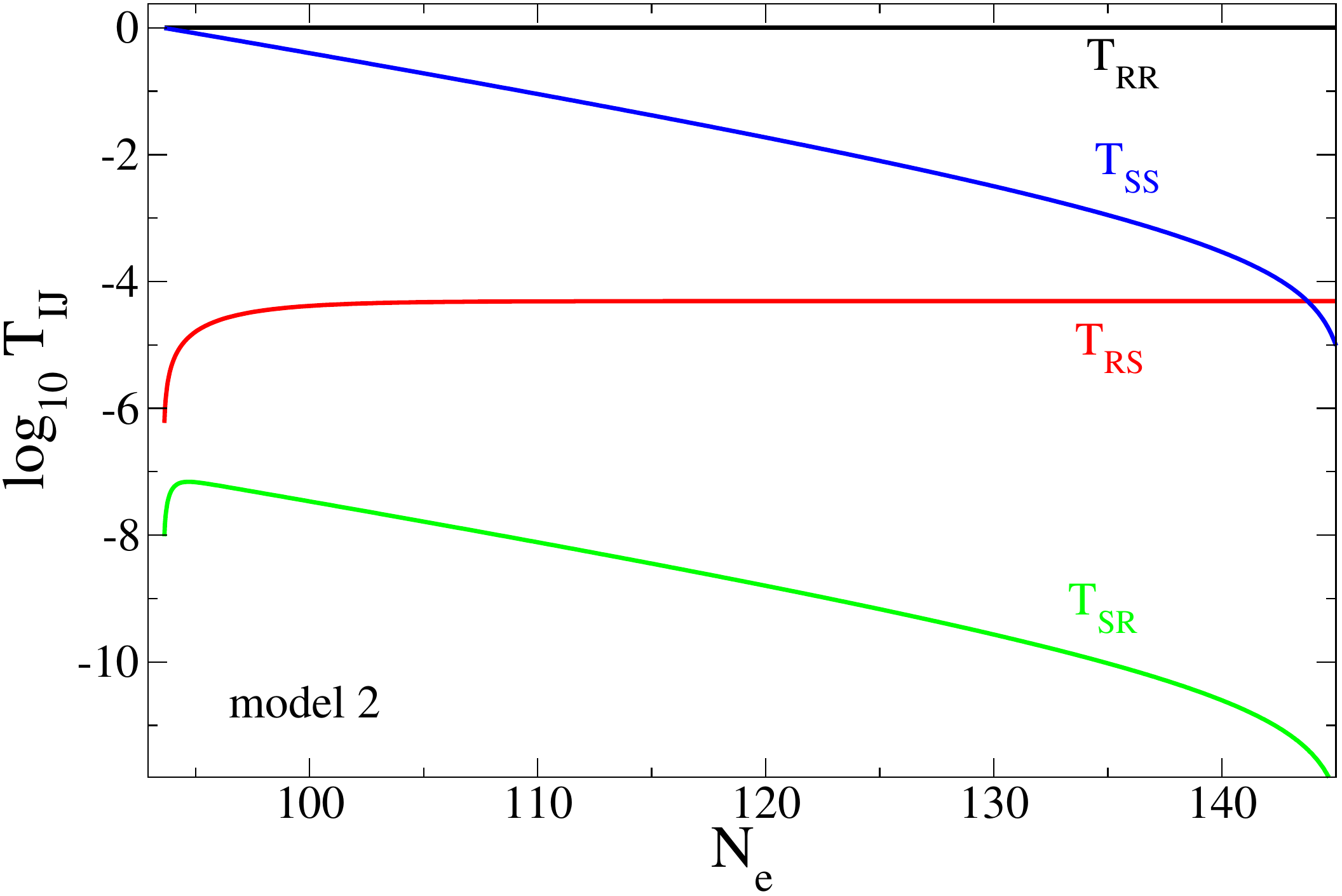}
 \caption{Evolution of transfer matrix elements for the adiabatic and
isocurvature perturbations, versus number of $e$-foldings, for models
1 and 2 from table \ref{tab1}.}
 \label{fig:trs}
\end{center} 
\end{figure*}

\subsection{Isocurvature fluctuations}
During inflation, the components of the canonically normalized fields
$U,V$ can fluctuate by order $H/(2\pi)$, where $H$ is the Hubble
parameter. Fluctuations $d\sigma$ normal to the inflaton trajectory are entropy
modes, and they could become observable isocurvature fluctuations if
they decay into different species than the adiabatic fluctuations,
that are parallel to the trajectory.  The relation between
adiabatic/entropy perturbations and the canonical field fluctuations
is given in eq.\ (\ref{dseq}).

To find the observable entropy fluctuations, we need to compare 
$(d\sigma,ds)$ to the directions in field space that correspond to baryon
number fluctuations $dB$, and the orthogonal direction, that will
be related to $(dU,dV)$ through some different rotation angle $\beta$.
Numerically we find that $\beta\cong 0$ during inflation, implying
that the entropy perturbations are purely in the baryon number
(compensated by radiation) to a good
approximation, known as BDI (baryon density isocurvature). This can be seen starting from the definition of
baryon density from the zeroth component of the baryon current 
carried by $\phi$,
\be
	n_B = j_B^0 = -2i(\phi\dot\phi^* - \phi^*\dot\phi) = 
	2(Y\dot X - X\dot Y)
\label{nBeq}
\ee
leading to the fluctuation
\be
	\delta n_B = 2(\dot X\,\delta Y - \dot Y\,\delta X) + \dots
\label{dnBeq}
\ee
where the omitted terms are subleading in the slow roll approximation.
The direction of the fluctuation (\ref{dnBeq}) turns out 
numerically to be 
very nearly orthogonal to the inflaton trajectory in field space.
Although both $\sigma$ and $s$ decay into quarks during reheating,
only $s$ decays encode the baryon asymmetry, whereas $\sigma$ decays
equally into quarks and antiquarks, that thermalize with the rest of
the SM degrees of freedom.

We closely follow the formalism of ref.\ \cite{Byrnes:2006fr} (see
also ref.\ \cite{Gordon:2000hv}) to compute the power in isocurvature.
The main task is to numerically solve the equations for the evolution
of the perturbations $dU,\,dV$ 
between horizon crossing and the
end of inflation,
\bea
	dU'' &=& -C_1 dU' -3(\bar\eta_{UU} dU + \bar\eta_{UV} dV)
	+ U'dC \nn\\ &+& (U'^2)' dU + (U' V')' dV\nn\\
	dV'' &=& -C_1 dV' -3(\bar\eta_{VV} dV + \bar\eta_{UV} dU)
	+ V'dC \nn\\ &+& (V'^2)' dV + (U' V')' dU
\label{fluc_eqs}
\eea
and to relate them to the adiabatic/isocurvature perturbations
$d\sigma,\,ds$ using eq.\ \ref{dseq}).  Here primes denote
$d/dN_e$, $C_1 = 3 + H'/H$, and $dC = C_1(U'dU + V'dV)$.
The barred parameters $\bar\eta_{ij}$ are defined as in eq.\ 
(\ref{srp}), except that we divide by the total energy density $\rho = 3
H^2$
instead of $V_E$, so that the equations remain valid even when the
slow-roll approximation is not.

The transfer function for the curvature (adiabatic) and entropy perturbations
is a matrix 
\be
	\left({T_{RR}\atop T_{SR}}\,{T_{RS}\atop T_{SS}}\right)
\ee
that relates the amplitudes of $(d\sigma,ds)$ at horizon
crossing to those at a later time, after inflation.  We can get the
matrix elements by solving the system (\ref{fluc_eqs}) from the
respective initial conditions $(d\sigma,ds)=(1,0)$ and $(0,1)$.
The results are shown for the two benchmark models in fig.\
\ref{fig:trs}.   The adiabatic perturbation is conserved, resulting
in $T_{RR}=1$, and the $T_{SR}$ element is always very small, in
accordance with the slow-roll prediction $T_{SR} = 0$ 
\cite{Byrnes:2006fr}, meaning
that there is negligible conversion of entropy to adiabatic modes.

For all cases in our MCMC, the entropy autocorrelation $T_{SS}\ll 0.1$ is
too small to be observable, but in some cases like in model 1, the
cross-correlation $T_{RS}$ is significant.  It is related to the
correlation angle defined by
\be
	\cos\Delta = {T_{RS}\over \sqrt{1+T_{RS}^2}}
\ee
which is constrained by Planck as $|\cos\Delta|\lesssim
0.1$-$0.3$, depending upon pivot scale $k_*$ and which 
datasets are combined.  (Ref.\ \cite{Akrami:2018odb} notes that
the constraints on BDI correlation are the same as for cold dark
matter isocurvature, CDI.)
Therefore model 1 is an example where the predicted BDI correlation
is close to the experimental sensitivity.

\begin{figure}[t]
\begin{center}
 \includegraphics[scale=0.4]{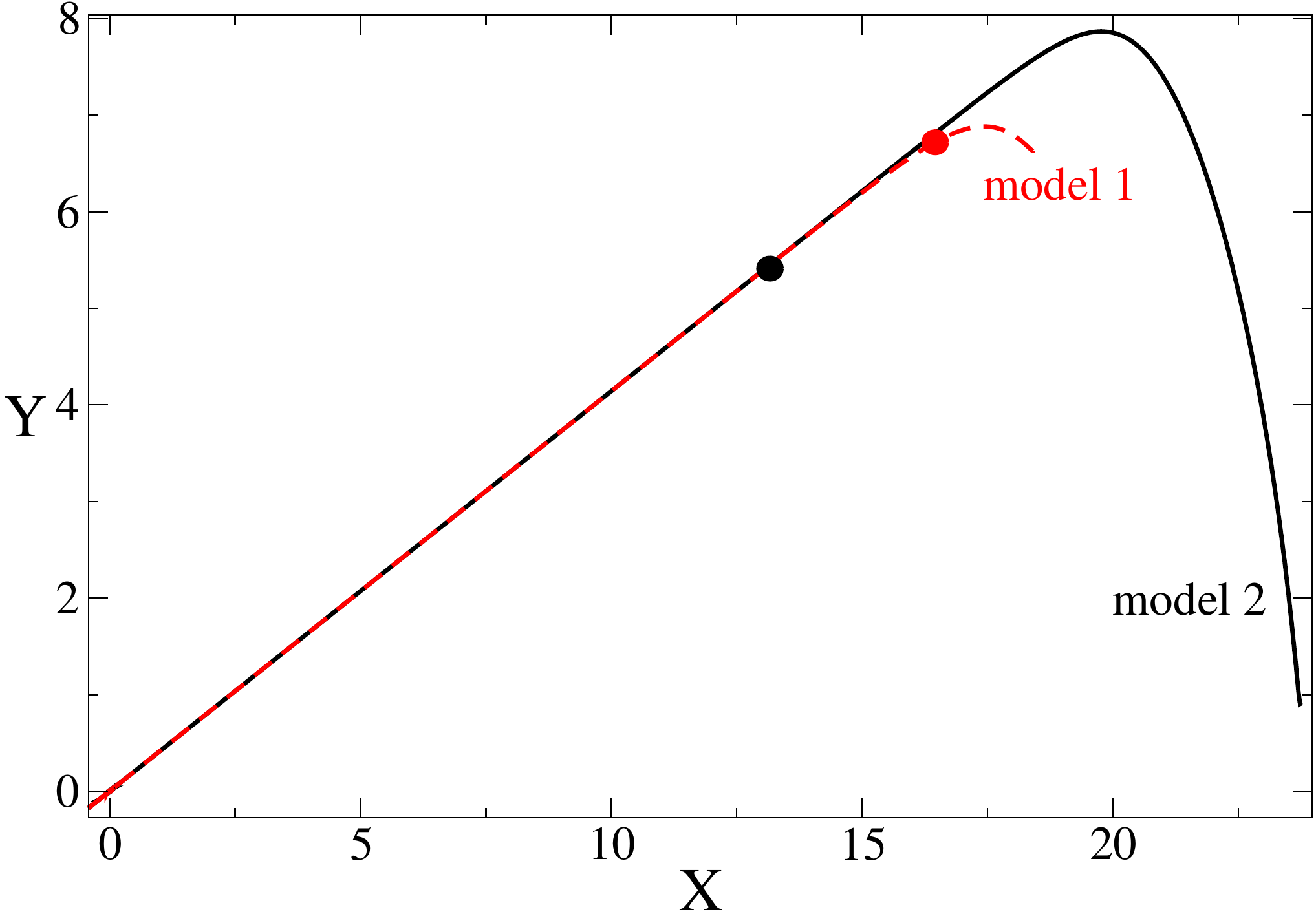}
 \caption{Inflaton trajectories in field space for the benchmark
models.  Horizon crossing is indicated by the heavy dot.}
 \label{fig:traj}
\end{center} 
\end{figure}

\begin{figure}[t]
\begin{center}
 \includegraphics[scale=0.4]{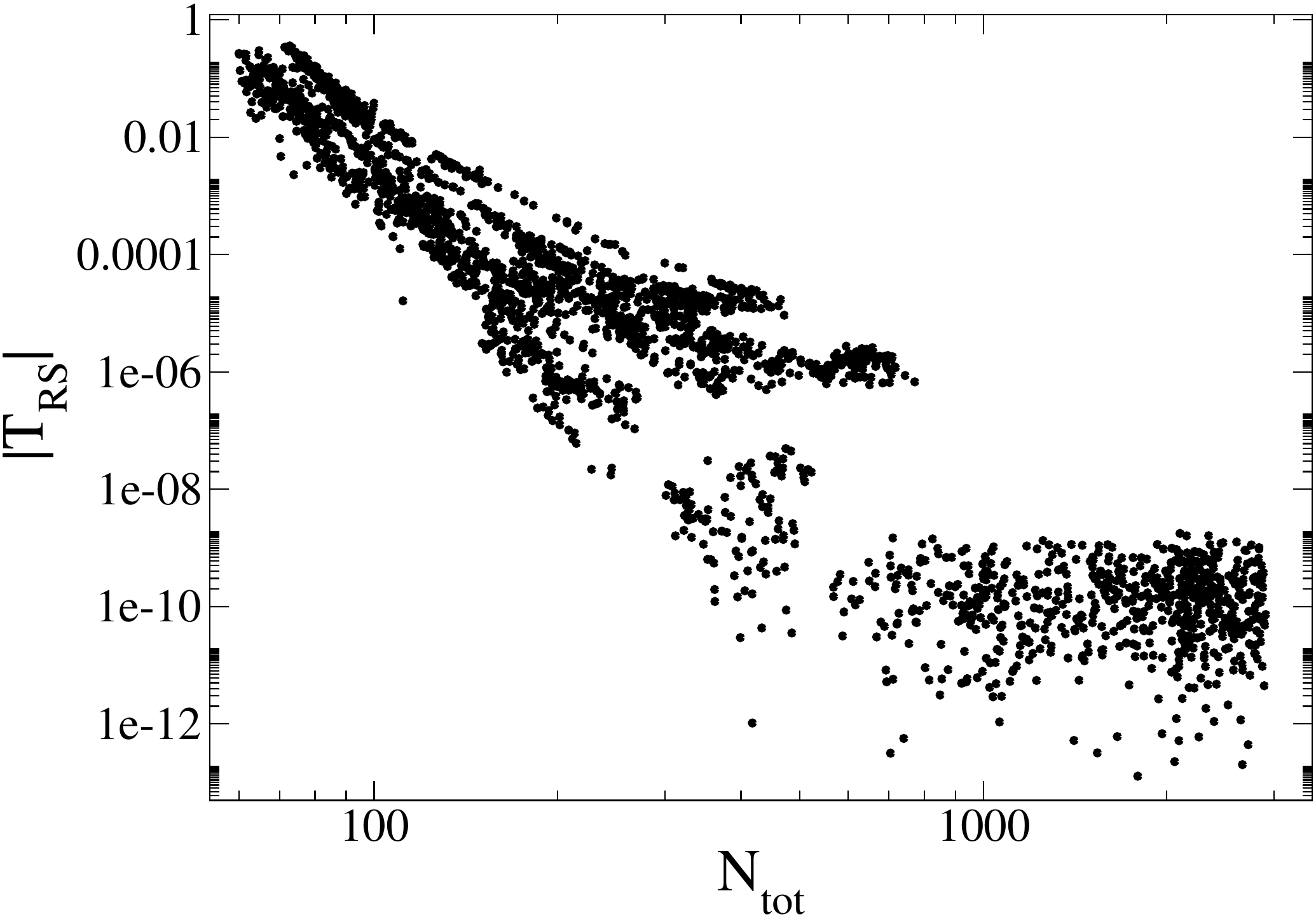}
 \caption{Scatter plot of isocurvature correlation $|T_{RS}|$ versus
the total number of $e$-foldings of inflation $N_{\rm tot}$ from the
MCMC.}
 \label{fig:trs-ntot}
\end{center} 
\end{figure}

The models with large BDI require somewhat special initial conditions,
in which the total duration of inflation is not more than $\sim 80$\
$e$-foldings.  This is because significant curvature of the inflaton
path in field space is needed during horizon crossing for generating
isocurvature.  Models with long periods of inflation tend to have 
such curvature earlier than horizon crossing, subsequently becoming
nearly linear and thus resembling single-field inflation.  This
is illustrated for the two benchmark models in fig.\ \ref{fig:traj},
that shows the field trajectories and horizon crossing points.
It is further borne out by fig.\ \ref{fig:trs-ntot}, showing the
correlation between $|T_{RS}|$ and total number of $e$-foldings $N_{\rm
tot}$ for models within an MCMC chain satisfying $\chi^2 < 10$. 
On the other hand, models like our benchmark model 2, having longer
periods of inflation, lead to predictions that are
relatively insensitive to the initial conditions, since the field
trajectory settles into a unique trough in the potential.

\subsection{Baryogenesis and reheating}
\label{bgrh}

\begin{figure}[t]
\begin{center}
 \includegraphics[scale=0.4]{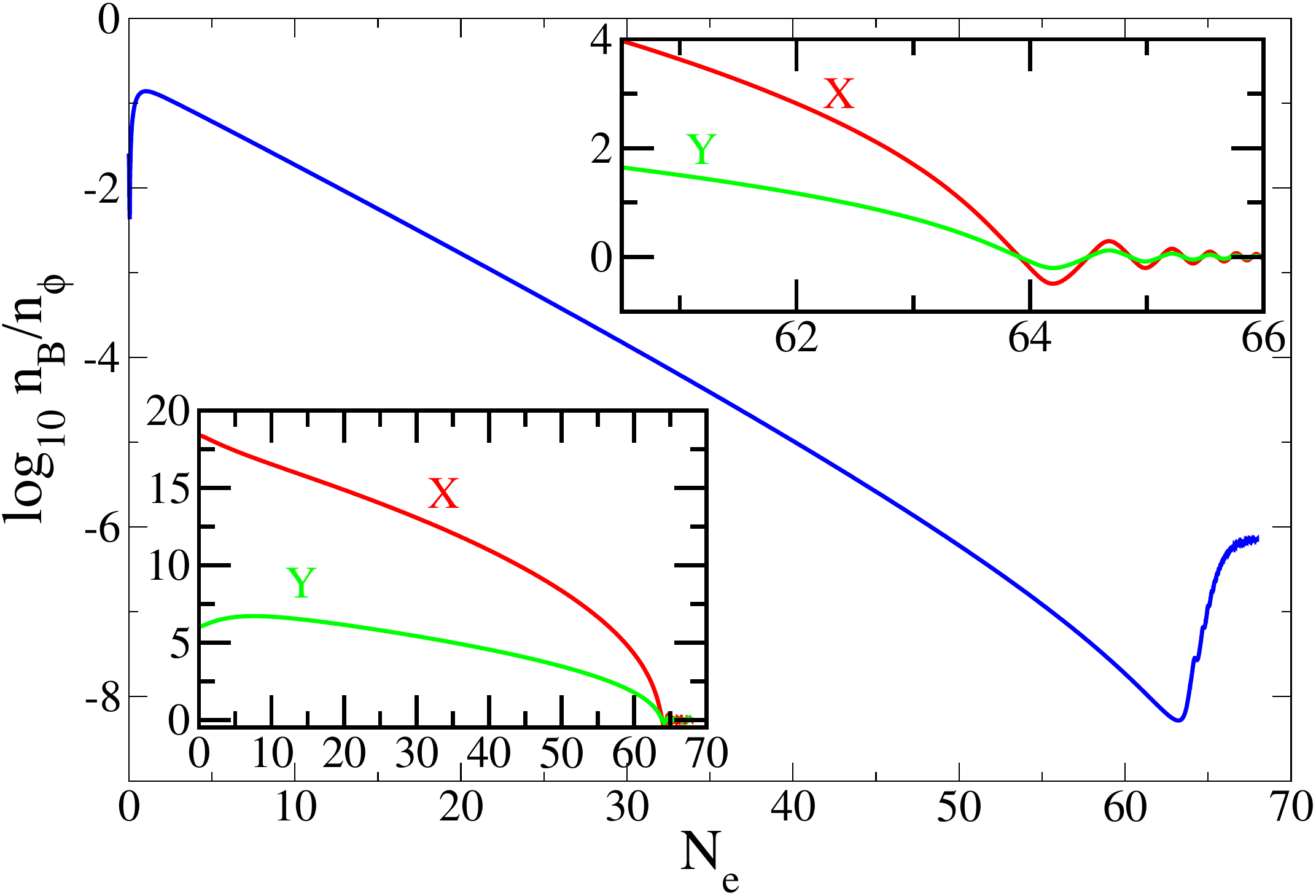}
 \caption{Baryon to inflaton ratio during inflation and shortly
after its end, versus number of $e$-foldings $N_e$, for benchmark
model 1.  Insets show the evolution of the field components,
$\phi = (X+iY)/\sqrt{2}$.}
 \label{fig:bau1}
\end{center} 
\end{figure}

To compute the baryon asymmetry, we use the baryon density stored in
$\phi$, eq.\ (\ref{nBeq}).
It is convenient to compare this to the number density of $\phi$
particles, prior to reheating,
\be
	n_\phi = {\rho_\phi\over m_\phi}
\ee
since the ratio $\eta = n_B/n_\phi$ reaches a constant value that
we denote as $\eta_e$
at the end of
inflation, during the period of $\phi$ oscillations around the minimum
of the potential.   The time evolution of $\eta$ is illustrated for
model 1 in fig.\ \ref{fig:bau1}.

\begin{figure}[t]
\begin{center}
 \includegraphics[scale=0.4]{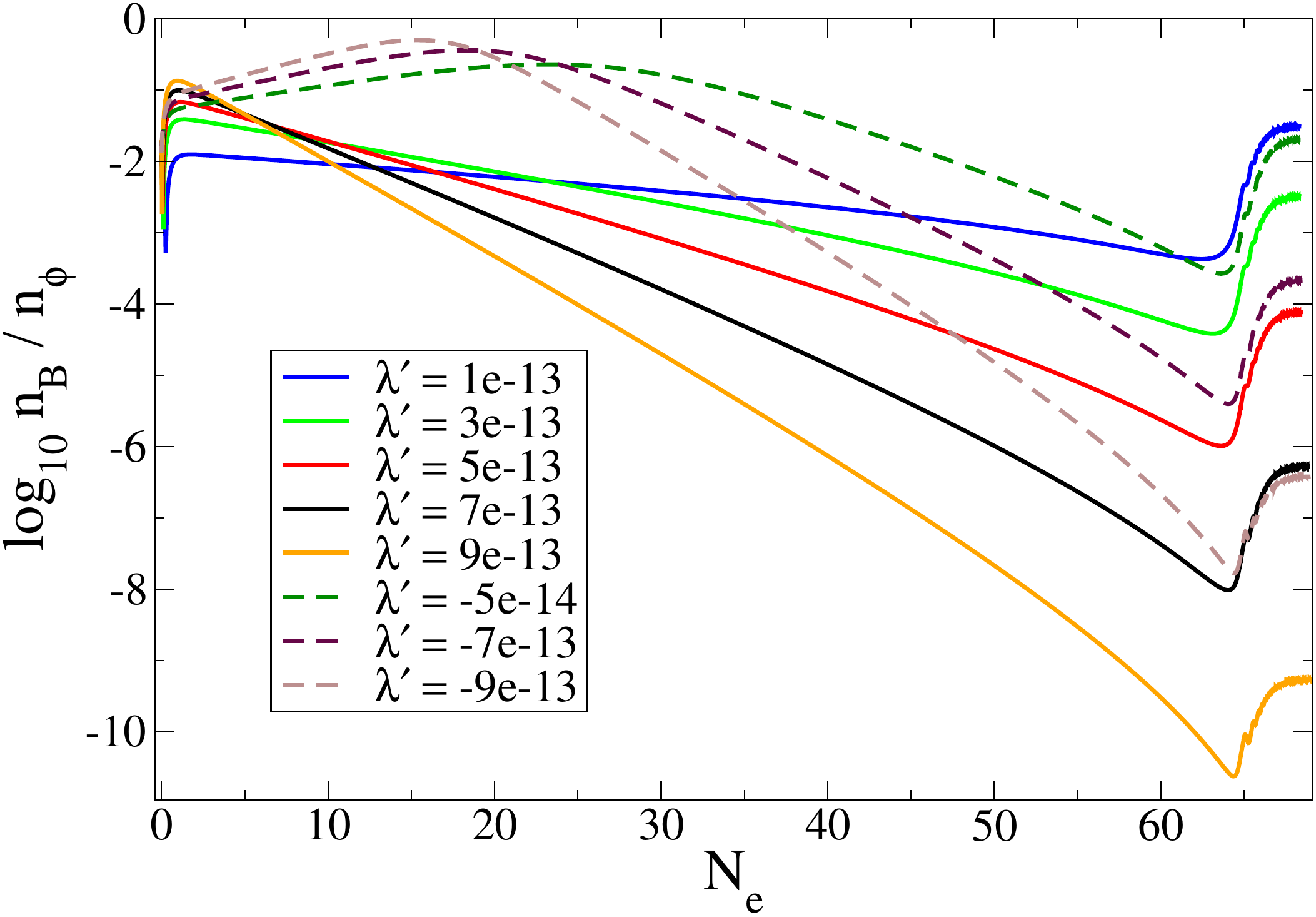}
 \caption{Baryon to inflaton ratio during inflation and shortly
after its end, versus number of $e$-foldings $N_e$, for several values
of $\lambda'$.  Other potential parameters are fixed at those
of model 1.  The curves are in the same order as the key, from top to
bottom at late times.  Positive values of $\lambda'$ are shown with
solid curves, negative with dashed.}
 \label{fig:bau}
\end{center} 
\end{figure}

Reheating occurs at the time $t_{\rm rh} = 1/\Gamma_\phi$ where $\Gamma_\phi$
is the decay width of $\phi$.  Defining $n_\phi = n_{\phi,e}$ at the
end of inflation ($t=t_e$), $n_\phi$ at the time of reheating will be
\bea
	n_{\phi,\rm rh} &=& n_{\phi,e}\left(a_e\over a_{\rm rh}\right)^3 = 
	{n_{\phi,e}\over 
	\left(1 + \frac{3}{2}\sqrt{m_\phi n_{\phi,e}\over 3
m_P^2}(t_{\rm rh}-t_e)\right)^2}\nn\\
	&\cong& {4m_P^2\,\Gamma_\phi^2\over 3m_\phi} 
\label{rho_rh}
\eea
where we used the fact that the $\phi$ oscillations matter-dominate
the universe until reheating, and $t_{\rm rh}\gg t_e$.  The value of $n_{\phi,\rm rh}$
is independent of $n_{\phi,e}$, so long as the latter is large enough to provide 
sufficient expansion of the universe prior to reheating.  This will
be true if the energy density at the end of inflation 
is much greater than that at reheating.

The baryon-to-entropy ratio at reheating is given by
\be
	\eta_B = \eta_e {n_{\phi,\rm rh}\over s}
\label{etaeeq}
\ee
with $s = (2\pi^2/45)g_* T_{\rm rh}^3$ and reheat temperature 
\cite{Kofman:1997yn}
\bea
	T_{\rm rh} &=& \left(90\over
\pi^2\,g_*\right)^{1/4}\left(\Gamma_\phi\, m_P\right)^{1/2}\nn\\
	&=& 1.7\times 10^{14}\,{\rm GeV} \left(\lambda''\over
10^{-2}\right)\left( m_\phi/m_P\over 5\times 10^{-7}\right)^{1/2}
\label{rhTemp}
\eea
Including a factor of $36/111$ \cite{Harvey:1990qw}
for the reduction of baryon number by redistribution into lepton
number by sphalerons, it follows that
\be
	\eta_B \cong 6.1\times 10^{-4}\,\eta_e\,\lambda'' 
	\left(m_P\over m_\phi\right)^{1/2}
\label{etaBeq}
\ee
which is conserved into the late universe.
The measured value is
$\eta_B = 8.6\times 10^{-11}$  \cite{Aghanim:2018eyx}.

The coupling $\lambda''$ should be
small in order to justify the perturbative reheating assumption, 
but from the point of view of technical naturalness, it need not be
very small.  A three-loop diagram involving $\lambda''$ renormalizes
the $\lambda|\phi|^4$ interaction, giving the estimate
\be
	\lambda'' \lesssim (16\pi^2)^{3/4}(\lambda/36)^{1/4} \cong
0.05
\label{lpbound}
\ee
to avoid destabilizing the inflationary potential by quantum
corrections.

\begin{figure}[t]
\begin{center}
 \includegraphics[scale=0.4]{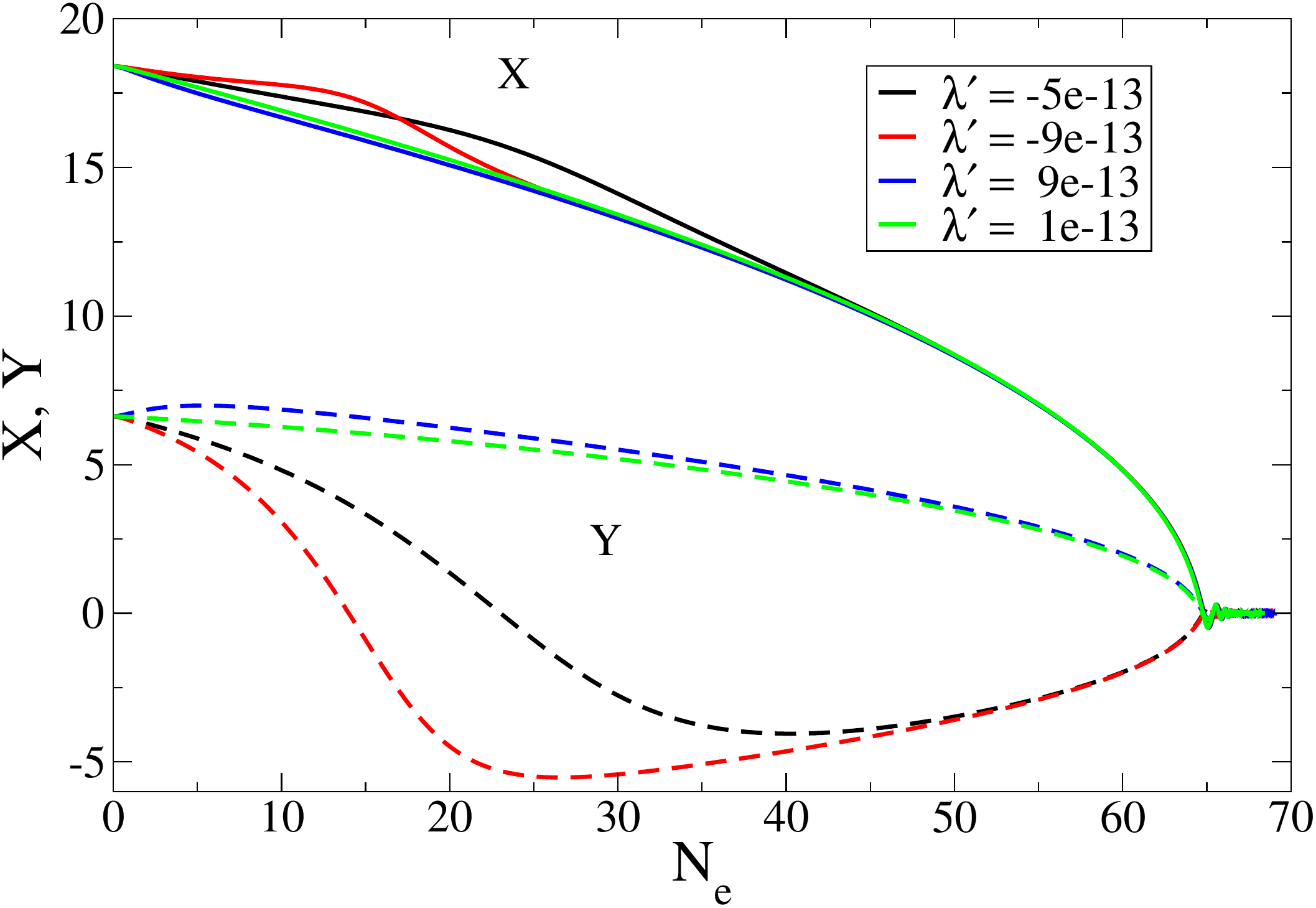}
 \caption{Inflaton trajectories $X(N_e)$ (solid) and $Y(N_e)$ (dashed)
for four different values of the baryon-violating coupling $\lambda'$.
Other parameters are fixed to those of model 1.}
 \label{fig:traj-lp}
\end{center} 
\end{figure}

The baryon asymmetry generated during inflation depends sensitively on
the value of the $B$-violating coupling $\lambda'$.  In fig.\
\ref{fig:bau} we again show how $\eta_B$ evolves with $N_e$ from the
beginning of inflation until shortly after it
ends, but for a range of different values of the baryon-violating
coupling $\lambda'$.   The effect of the $\phi$ oscillations can be seen briefly
around $N_e=65$, but these are quickly Hubble-damped and $\eta_B$
settles to a constant value that we have identified as $\eta_e$ in
eq.\ (\ref{etaeeq}).  The dependence of the final baryon asymmetry
is not monotonic.  At first this may be surprising, since one can
derive the time-dependence of $n_B$ from the inflaton field equations,
\bea
	\dot n_B &=& 2i(\phi^*\ddot\phi - \ddot\phi^*\phi)\nn\\
	&=& - 3H n_B + 4\lambda'(\phi^4+\phi^{*4})
\eea
However one finds that $\lambda'$ has an important effect on the
background inflaton trajectory, which explains the nonlinear
dependence.
This is illustrated in fig.\ \ref{fig:traj-lp}.
Hence the processes of inflation and baryogenesis are 
nontrivially intertwined in our model: adjusting $\lambda'$ can affect
not only $\eta_B$ but also the inflationary observables.

The effects of $B$ violation after inflation are negligible.
At low energies, integrating out $\phi$ and $\Phi_i$ leads to 
a dimension-36 operator involving 24 quarks.  It could induce
conversion of four neutrons into their antiparticles in a neutron
star, but the rate is far too small to be significant.  In the early
universe, we must check that the $\Delta B = 8$, $\Phi_i^{12}$ operator 
induced by $\phi$ exchange is out of equilibrium, to avoid washing out
the $B$ asymmetry.  The rate can be estimated as
\be
	\Gamma_{\Delta B = 8} \sim {{\lambda'}^2 {\lambda''}^8 T^{17}
	\over m_\phi^{16}} \lesssim {T^2\over m_P}
\ee
By demanding that the decoupling temperature exceed the reheat
temperature $T_{\rm rh}$ in eq.\ (\ref{rhTemp}), we find a constraint
\be
	\lambda'' \lesssim 20\, {m_\phi^{17/46}\over
	(\lambda')^{2/23}} \sim 1.5
\ee
which is more lenient than the consistency requirement
(\ref{lpbound}).

\begin{table*}[]
\centering
\tabcolsep 2.5pt
\begin{tabular}{|c|c|c|c|c|}
\hline
system & $K^0$-$\bar K^0$ & $K^0$-$\bar K^0$ & $B^0$-$\bar B^0$ &
$B_s^0$-$\bar B_s^0$ \\
\hline
coefficient & ${m_{\Phi_3}\,{\rm Re}[y_{3,dd}\, y^*_{3,ss}]^{-1/2}}$ &
${m_{\Phi_3}\,{\rm Im}[y_{3,dd}\, y^*_{3,ss}]^{-1/2}}$ &
${m_{\Phi_3}\,|y_{3,dd}\, y^*_{3,bb}|^{-1/2}}$ &
${m_{\Phi_3}\,|y_{3,ss}\, y^*_{3,bb}|^{-1/2}}$ \\
\hline
limit &  $1.1\times 10^3$\,TeV & $2.1\times 10^4$\,TeV & 
$990$\,TeV & $245$\,TeV \\

\hline   
\end{tabular}
\caption{Lower limits from meson-antimeson mixing on parameters entering the
Wilson coefficients of four-quark operators from integrating out the
heavy color triplet $\Phi_3$.
\label{tab2}}
\end{table*}

\section{Particle physics implications}

The colored scalars $\Phi_i$ can have observable effects at low energies.
If sufficiently light, they can be pair-produced at LHC.  The Yukawa
interactions in eq.\ (\ref{Phic}) have the same form as $R$-parity
violating coupling of squarks to quarks in supersymmetric models,
leading to various mass exclusions in the range $80$-$525$\,GeV
\cite{Sirunyan:2018rlj} or $100$-$600$\,GeV \cite{Aaboud:2017nmi}, 
depending upon the flavor structure of the couplings.

However heavier colored scalars can be probed indirectly, using
an effective field theory description where they are integrated out
to give dimension-6, four-quark operators.  For baryogenesis, the 
flavor structure of the new Yukawa couplings was not important, but
at low energies it can have an observable effect on the angular
distributions of jets at LHC, or flavor-changing neutral-currents
like meson-antimeson oscillations.  Using chiral Fierz identities
\cite{Nishi:2004st},  the effective Lagrangian is
\bea
{\cal L} &=& -	\sum_{A=1,2}  \delta^{ae}_{bd}\,{y_{A,ij} y^*_{A,kl}
	\over 2 m_{\Phi_A}^2}
	(\bar u_i^a \gamma^\mu P_\R u_{k,b})(\bar d_j^e 
	\gamma_\mu P_\R d_{l,d})\nn \\
   &-& {y_{3,ii}y^*_{3,jj} \over  m_{\Phi_3}^2}
(\bar d_i^a \gamma^\mu P_\R d_{j,a})(\bar d_i^b 
	\gamma_\mu P_\R d_{j,b})
 \eea
where $a,b,d,e$ are color indices, $i,j,k,l$ label flavor, and $P_\R$
projects onto right-handed chirality.  In the bottom line we have
specialized to the case where $i\neq j$ and the operator contributes to
meson-antimeson oscillations, since these combinations are much more
severely constrained than the flavor-diagonal ones, or those connecting
mesons of different masses.

From dijet angular distributions, CMS finds a limit of \cite{Sirunyan:2018wcm}
\be
	\left( m_{\Phi}^2\over y y^*\right)^{1/2} \gtrsim 7\,{\rm TeV}
\ee
for flavor-diagonal operators, presumably of the first generation (since
the limit on higher generation quarks will be somewhat weakened by parton
distribution functions).  However $K^0$-$\bar K^0$, $B^0$-$\bar B^0$
and $B_s^0$-$\bar B_s^0$ mixing give more stringent constraints
\cite{Bona:2017cxr}, shown in table \ref{tab2}.

\section{Conclusions}

We have studied a new model of inflation with the novel feature that the
inflaton carries baryon number, and it can produce the baryon asymmetry
via the Affleck-Dine mechanism, mostly during inflation, with relatively
small evolution over the few $e$-foldings after inflation ends. It is a
simple but complete model, including  a calculable
perturbative reheating mechanism that allows one to make definite
predictions for the inflationary observables, given a set of input
parameters.  One testable prediction is that the tensor-to-scalar ratio
$r$ is likely to be observable, depending upon the value of the
nonminimal coupling of the inflaton to gravity.  For the values $\xi \sim
0.01-1$ considered in this work, we find $r>0.04$, which is within the
sensitivity of upcoming CMB experiments.  For example LiteBIRD will probe
values down to $r\sim 10^{-3}$ \cite{Hazumi:2019lys}.

Since ours is a two-field inflation model, another possible signal is
correlated baryon isocurvature-adiabatic fluctuations that have been
constrained by the Planck collaboration.  We find that these can occur
at an observable level if the total duration of inflation did not greatly
exceed the canonical minimum number of $e$-foldings, $N_{\rm tot}\sim 60$.  In this
case the inflaton trajectory can turn significantly in field space around the
time of horizon crossing.  We are not aware of other models in the
literature that predict baryon isocurvature perturbations.

The model relies upon new colored scalar particles in order to
transfer the baryon asymmetry from the inflaton to the standard model quarks.  These could
have observable effects in laboratory experiments if sufficiently light,
even at the scale of $10^4$\,TeV for $K^0$-$\bar K^0$ oscillations.
The colored scalars could also mediate purely hadronic
rare flavor changing decays, that we have not considered here.  The new
source of baryon violation needed for baryogenesis is however hidden at
the high scale the inflaton mass $\sim 10^{-7}\,m_P$, out of reach of
laboratory probes.

We have considered only the simplest scenario for reheating.  It is
possible that sufficiently large values of $\lambda''$ could lead to
more efficient reheating through parametric resonance
\cite{Kofman:1997yn}.  To our knowledge, this has not been previously
studied for couplings of the form $\phi\Phi^3$ such as are present
in our model.  Moreover we ignored the Higgs portal coupling 
$|\phi|^2|H|^2$ which could reduce the baryon asymmetry by producing
extra radiation.  We leave these issues for future study.

\smallskip
{\bf Acknowledgment.}  We thank C.\ Byrnes and R.\ Namba 
for very helpful correpondence or discussions.
Our work is supported by NSERC (Natural Sciences and Engineering
Research Council, Canada).  MP is supported by the Arthur B.\ McDonald
Institute for Canadian astroparticle physics research.\\

\smallskip {\bf Note added.}  After finishing this work, we became aware
of refs.\ \cite{Hertzberg:2013mba,Takeda:2014eoa} which study the same 
idea, but do
not address the problem of chaotic inflation models being ruled out by
current CMB data.  These works do not consider 
renormalizable operators for the
baryon asymmetry transfer and reheating, and ref.\ 
\cite{Hertzberg:2013mba} has different
predictions for the generated isocurvature (not considered by ref.\
\cite{Takeda:2014eoa}).


\begin{thebibliography}{10}

\bibitem{Affleck:1984fy} 
  I.~Affleck and M.~Dine,
  ``A New Mechanism for Baryogenesis,''
  Nucl.\ Phys.\ B {\bf 249}, 361 (1985).
  doi:10.1016/0550-3213(85)90021-5

\bibitem{Sakharov:1967dj} 
  A.~D.~Sakharov,
  ``Violation of CP Invariance, C asymmetry, and baryon asymmetry of the universe,''
  Pisma Zh.\ Eksp.\ Teor.\ Fiz.\  {\bf 5}, 32 (1967)
  [JETP Lett.\  {\bf 5}, 24 (1967)]
  [Sov.\ Phys.\ Usp.\  {\bf 34}, no. 5, 392 (1991)]
  [Usp.\ Fiz.\ Nauk {\bf 161}, no. 5, 61 (1991)].
  doi:10.1070/PU1991v034n05ABEH002497

\bibitem{Linde:1983gd} 
  A.~D.~Linde,
  ``Chaotic Inflation,''
  Phys.\ Lett.\  {\bf 129B}, 177 (1983).
  doi:10.1016/0370-2693(83)90837-7

\bibitem{Akrami:2018odb} 
  Y.~Akrami {\it et al.} [Planck Collaboration],
  ``Planck 2018 results. X. Constraints on inflation,''
  arXiv:1807.06211 [astro-ph.CO].

\bibitem{Okada:2010jf} 
  N.~Okada, M.~U.~Rehman and Q.~Shafi,
  ``Tensor to Scalar Ratio in Non-Minimal $\phi^4$ Inflation,''
  Phys.\ Rev.\ D {\bf 82}, 043502 (2010)
  doi:10.1103/PhysRevD.82.043502
  [arXiv:1005.5161 [hep-ph]].

\bibitem{Linde:2011nh} 
  A.~Linde, M.~Noorbala and A.~Westphal,
  ``Observational consequences of chaotic inflation with nonminimal coupling to gravity,''
  JCAP {\bf 1103}, 013 (2011)
  doi:10.1088/1475-7516/2011/03/013
  [arXiv:1101.2652 [hep-th]].

\bibitem{Evans:2015mta} 
  J.~L.~Evans, T.~Gherghetta and M.~Peloso,
  ``Affleck-Dine Sneutrino Inflation,''
  Phys.\ Rev.\ D {\bf 92}, no. 2, 021303 (2015)
  doi:10.1103/PhysRevD.92.021303
  [arXiv:1501.06560 [hep-ph]].

\bibitem{Cline:2006hu} 
  J.~M.~Cline,
  ``String Cosmology,''
  hep-th/0612129.

\bibitem{Byrnes:2006fr} 
  C.~T.~Byrnes and D.~Wands,
  ``Curvature and isocurvature perturbations from two-field inflation in a slow-roll expansion,''
  Phys.\ Rev.\ D {\bf 74}, 043529 (2006)
  doi:10.1103/PhysRevD.74.043529
  [astro-ph/0605679].

\bibitem{Gordon:2000hv} 
  C.~Gordon, D.~Wands, B.~A.~Bassett and R.~Maartens,
  ``Adiabatic and entropy perturbations from inflation,''
  Phys.\ Rev.\ D {\bf 63}, 023506 (2001)
  doi:10.1103/PhysRevD.63.023506
  [astro-ph/0009131].



\bibitem{Liddle:2003as} 
  A.~R.~Liddle and S.~M.~Leach,
  ``How long before the end of inflation were observable perturbations produced?,''
  Phys.\ Rev.\ D {\bf 68}, 103503 (2003)
  doi:10.1103/PhysRevD.68.103503
  [astro-ph/0305263].

\bibitem{Kofman:1997yn} 
  L.~Kofman, A.~D.~Linde and A.~A.~Starobinsky,
  ``Towards the theory of reheating after inflation,''
  Phys.\ Rev.\ D {\bf 56}, 3258 (1997)
  doi:10.1103/PhysRevD.56.3258
  [hep-ph/9704452].

\bibitem{Harvey:1990qw} 
  J.~A.~Harvey and M.~S.~Turner,
  ``Cosmological baryon and lepton number in the presence of electroweak fermion number violation,''
  Phys.\ Rev.\ D {\bf 42}, 3344 (1990).
  doi:10.1103/PhysRevD.42.3344

\bibitem{Aghanim:2018eyx} 
  N.~Aghanim {\it et al.} [Planck Collaboration],
  ``Planck 2018 results. VI. Cosmological parameters,''
  arXiv:1807.06209 [astro-ph.CO].

\bibitem{Sirunyan:2018rlj} 
  A.~M.~Sirunyan {\it et al.} [CMS Collaboration],
  ``Search for pair-produced resonances decaying to quark pairs in proton-proton collisions at $\sqrt{s}=$ 13 TeV,''
  Phys.\ Rev.\ D {\bf 98}, no. 11, 112014 (2018)
  doi:10.1103/PhysRevD.98.112014
  [arXiv:1808.03124 [hep-ex]].

\bibitem{Aaboud:2017nmi} 
  M.~Aaboud {\it et al.} [ATLAS Collaboration],
  ``A search for pair-produced resonances in four-jet final states at $\sqrt{s} =$ 13 TeV with the ATLAS detector,''
  Eur.\ Phys.\ J.\ C {\bf 78}, no. 3, 250 (2018)
  doi:10.1140/epjc/s10052-018-5693-4
  [arXiv:1710.07171 [hep-ex]].

\bibitem{Nishi:2004st} 
  C.~C.~Nishi,
  ``Simple derivation of general Fierz-like identities,''
  Am.\ J.\ Phys.\  {\bf 73}, 1160 (2005)
  doi:10.1119/1.2074087
  [hep-ph/0412245].


\bibitem{Sirunyan:2018wcm} 
  A.~M.~Sirunyan {\it et al.} [CMS Collaboration],
  ``Search for new physics in dijet angular distributions using proton–proton collisions at $\sqrt{s}=$ 13 TeV and constraints on dark matter and other models,''
  Eur.\ Phys.\ J.\ C {\bf 78}, no. 9, 789 (2018)
  doi:10.1140/epjc/s10052-018-6242-x
  [arXiv:1803.08030 [hep-ex]].

\bibitem{Bona:2017cxr} 
  M.~Bona [UTfit Collaboration],
  ``Latest results for the Unitary Triangle fit from the UTfit Collaboration,''
  PoS CKM {\bf 2016}, 096 (2017).
  doi:10.22323/1.291.0096

\bibitem{Hazumi:2019lys} 
  M.~Hazumi {\it et al.},
  ``LiteBIRD: A Satellite for the Studies of B-Mode Polarization and Inflation from Cosmic Background Radiation Detection,''
  J.\ Low.\ Temp.\ Phys.\  {\bf 194}, no. 5-6, 443 (2019).
  doi:10.1007/s10909-019-02150-5

\bibitem{Hertzberg:2013mba} 
  M.~P.~Hertzberg and J.~Karouby,
  ``Generating the Observed Baryon Asymmetry from the Inflaton Field,''
  Phys.\ Rev.\ D {\bf 89}, no. 6, 063523 (2014)
  doi:10.1103/PhysRevD.89.063523
  [arXiv:1309.0010 [hep-ph]].

\bibitem{Takeda:2014eoa} 
  N.~Takeda,
  ``Inflatonic baryogenesis with large tensor mode,''
  Phys.\ Lett.\ B {\bf 746}, 368 (2015)
  doi:10.1016/j.physletb.2015.05.039
  [arXiv:1405.1959 [astro-ph.CO]].

\end{thebibliography}
\end{document}